# THE OPTIMAL FORM OF DISTRIBUTION NETWORKS APPLIED TO THE KIDNEY AND LUNG

WALTON R. GUTIERREZ
*Touro College, 27 West 23rd Street, New York, NY 10010*
*waltong@touro.edu*

**Abstract**

A model is proposed to minimize the total volume of the main distribution networks of fluids in relation to the organ form. The minimization analysis shows that the overall exterior form of distribution networks is a modified ellipsoid, a geometric form that is a good approximation to the external anatomy of the kidney and lung. The variational procedure implementing this minimization is similar to the traditional isoperimetric theorems of geometry.
A somewhat expanded version of this article in Section 4 will be published in the ***Journal of Biological Systems***, *World Scientific Publishing*.

*Keywords:* Network Volume, Fractal Tree Volume, Self-Similar Network, Lung and Kidney, Optimal Form, Isoperimetric Principle. Quantitative Anatomy.

## 1. INTRODUCTION

At the allometric level, there is similarity between the kidney and lung of mammals, when these organs are anatomically described in terms of the dimensions of the nephrons and alveoli,[1-3] here generally referred to as the sites.
The kidney and lung are notable for their capacity to process large amount of fluids (liquids and gases) in a small organ volume, thus indicating that the distribution networks of fluids of these organs have a geometric design of great efficiency. A basic requirement in achieving this efficiency is the minimization of the total volume of the networks distributing the fluids, so that a larger part of the volume of the organ is available to the sites where the actual processing of fluids is performed. Computer simulations of distribution networks of simplified box-like organs already show the large effect of the organ form in the total volume of an efficient distribution network[4].
The model developed here links general geometric and fractal characteristics of distribution networks to the external form of the network, from the microscopic sites to the macroscopic anatomical organ form. Clearly, the minimization of the network volume is subjected to many diverse requirements among which the following biophysical conditions are considered in this article:
(*I*). Total bulk volume of the organ, regardless of organ form.
(*II*). Inlet and outlet features of the networks of the organ (ureter, main bronchus, etc).
(*III*). Sharing of space among the organs within the thorax and abdomen and special anatomic features such as the diaphragm that are needed to assist the movement of materials through the organ.
It will be shown that conditions (*I*) and (*II*), together with some elements of (*III*), are sufficient to derive an approximation to the kidney form and to the overall exterior form of both lungs above the diaphragm. A full accounting of condition (*III*) requires a very complex set of calculations, a goal that is outside the scope of this paper.

The kidney serves here as a primary example for the concepts and techniques of the model. Since the total volume of a kidney is small, the kidney is only partially conditioned by (*III*). The kidney shape is simpler in the sense that it does not change shape as much as the lung does during the breathing cycle.

A common use of variational techniques is found in the following classic isoperimetric theorem: for a given volume value, the geometric form with the least area is the sphere, which is also the form containing the maximum volume for a given area. In this model it is similarly discovered that the extrema (minima) of network volume for a given organ volume is found in the modified ellipsoid organ form, which also corresponds to the extrema (maxima) of organ volume for a given network volume.

Section 4 discusses the more general fractal characteristics of the organ networks, which do not conform to strict self-similarity[2]. In this section, a simple model of a network supplying a "rectangular organ" shows that a large set of networks of variable fractal features has a single optimal exterior form. The "rectangular organ" model further supports and illustrates some of the main modeling decisions.

This work is a continuation of a method introduced in Gutierrez[2] for the calculation of the volume of distribution networks with fractal characteristics.

## 2. THE DISTRIBUTION NETWORK AND THE ORGAN FORM MODEL

The starting point of the main model is a formula (Eq. (8) of Gutierrez [2]) for the calculation of the volume of a network of fluid distribution,

$$V_{net} = (P/V)\int X(\mathbf{r})dV \equiv (P/V)I_X \qquad (1)$$

where $P = C_{01}[A_0 p_1 + A_s N_s (1 - p_1)]$; $C_{01}$ is a constant; $p_1$ is a parameter such $0 < p_1 < 1$; $A_0$ is the cross section area of the main branch of the network such as the renal artery; $A_s$ is the cross section area of a branch supplying a site, such as a nephron or alveolus; $N_s$ is the number of sites in the organ ($\sim 10^6$ in the human kidney); $V$ is the organ volume; and $I_X$ is the integral over the organ volume region of the circulation distance function $X(\mathbf{r})$. This is the distance inside the network between the position $\mathbf{r} = (r_x, r_y, r_z)$ of a site of the organ, to the average inlet position of the network. A simple fractal model is shown in Sec. 4 illustrating the use of Eq. (1).

Before going into the details of the model, a broad summary of the main argument is that $I_X$ (Eq.(1)) is the quantity to be minimized, using the organ form (surface) as a variable while leaving $P$ and $V$ constant in this process with the appropriate boundary conditions representing the volumetric coexistence of the organs. In a first approximation $P$ may also be a function of $V$, thus keeping a constant $V$ also makes $P$ constant. The simple model shown in the Sec. 4 also discusses these arguments.

To model circulatory organs such as the kidney and lung, we should consider the three main networks: the arterial, the venous, and the urinary in the kidney and the bronchial in the lung. Due to the similarities and geometric parallelism of the venous and arterial networks within these organs, these two networks are summarized into a single network. In this modeling approximation, the organ contains two major networks: *A* and *B*. In the kidney they are the urinary network (*A*) and the combined arterial and venous network (*B*) and in the lung they are the bronchial network (*A*) and the blood network (*B*). Thus by extending Eq. (1) to the total volume of network *A* and *B*,

$$V_{netAB} = (P_A I_{XA} + P_B I_{XB})/V \qquad (2)$$

To describe the various network geometric functions, a conventional spherical coordinate system is introduced: $(r, \theta, \varphi) = (r, \Omega)$. The origin of the coordinate is placed at the origin of network *A* and the origin of network *B* is placed at a position $c_B$ on the $+r_z$ axis (see Fig. 1). The inlet/outlet connections of the organ are in the region of the $-r_x$ axis. For what follows in this Sec., the simplest and non-trivial form of the circulation distance function $X(r, \theta, \varphi)$ is chosen as, $X_A = f_{A1} r$ and $X_B = f_{B1} r_B$, where $r_B$ is the straight line distance between $(r_x, r_y, r_z)$ and $(0, 0, c_B)$, (see Fig.1). Additional corrections to this basic choice[2] may be necessary in the future, in the form of, $X(r, \theta, \varphi) = f_1 r + f_2 r^\nu \chi(\theta, \varphi)$, where, $f_1 > 1, f_1 > f_2$. Section 4 shows that the choice $X = f_1 r$, is quite adequate for representing many fractal distribution networks that may approximate the organ networks. Also there is an allometric argument given in Gutierrez[2] to choose this as the leading order term for the circulation distance function.

The region of integration for $r_x > 0$ is exteriorly delimited by the organ form represented by the surface $R(\theta, \varphi)$, (Fig. 1) which will be determined as a solution of the equation resulting from the minimization of the network volume. Going from $r_x > 0$ to $r_x < 0$ the region of integration near the origin of the networks (Fig. 1) can include another region delimited by the $R_{in}(\theta, \varphi)$ surface needed to provide space to connect the organ to the major inlet or outlet vessels of body circulation (such as renal artery, vein and ureter), and some other organs in the case of the lung. The function $R_{in}$ is treated as a given surface that is a boundary condition for *R*, and in this way condition (*II*) and part of condition (*III*) of the introduction are accounted for. The union of $R_{in}$ and *R* surfaces represent the whole organ form, and using them in Eq. (2) the network volume is

$$V_{netAB} = V^{-1} \int_{\Omega'} \int_{R_{in}(\Omega)}^{R(\Omega)} (Q_A r + Q_B r_B) r^2 dr d\Omega \equiv V^{-1} I(R) \qquad (3)$$

where $Q_A = P_A f_{A1}$, $Q_B = P_B f_{B1}$, and $d\Omega = \sin\varphi d\varphi d\theta$. A basic trigonometric application to the distance $r_B$ shows that $r_B(r, \varphi) = (r^2 + c_B^2 - 2rc_B \cos\varphi)^{1/2}$. Also, using the same variables and functions, the volume of the organ is

$$V = V(R) = \int_{\Omega'} \int_{R_{in}(\Omega)}^{R(\Omega)} r^2 dr d\Omega \qquad (4)$$

To minimize the networks volume under the variations of the organ form $R(\Omega)$ the integral $I(R)$ (Eq. (3)) is regarded as a functional of $R(\Omega)$. To incorporate condition (*I*) of the introduction, the minimization is done for a given value of the organ volume $V_o = V(R)$. A Lagrange multiplier $\lambda$ is introduced to combine (3) and (4), and determine the extrema in relation to *R* of the auxiliary functional $V_{aux}(R)$

$$V_{aux}(R) = V_o^{-1}I(R) + \lambda(V_o - V(R)) \equiv \int_{\Omega'} F_{aux}(R)d\Omega + \lambda V_o \qquad (5)$$

where $F_{aux}(R) = \int_{R_{in}(\Omega)}^{R(\Omega)}[V_o^{-1}(Q_A r + Q_B r_B(r,\varphi)) - \lambda]r^2 dr$. Since $F_{aux}(R)$ contains no derivatives of $R$, the problem is reduced to equating to zero the partial derivative of $F_{aux}(R)$ in relation to $R$, which implies the following equation for $R$

$$\partial F_{aux}/\partial R = [(Q_A R + Q_B r_B(R,\varphi))/V_o - \lambda]R^2 = 0. \qquad (6)$$

Next we discuss the properties of the solution of this last Eq. This Eq. is a simple quadratic equation for R and the solution formula with the square root in the denominator is chosen. Rewriting Eq. (6) as $aR^2 + 2bR - 2c = 0$, we have
$R(\varphi) = 2c/\{b \pm [b^2 + 2ac]^{1/2}\}$, where the extra factors of 2 are introduced to avoid a later redefinition of parameters. In this case, $a = (Q_B^2 - Q_A^2)/\lambda Q_A V_o = (e_c/c_B) - (1/c_L)$; $b = 1 - e_c\cos\varphi$, and $c = (c_L - e_c c_B)/2$, where the new parameters are $c_L = \lambda V_o/Q_A$ and $e_c = c_B Q_B^2/\lambda Q_A V_o$. Notice that the four original positive parameters $Q_A$, $Q_B$, $V_o$, $c_B$, together with $\lambda$ are reduced to the three effective parameters for $R(\varphi)$ which are $c_B$, $c_L$, and $e_c$. In terms of dimensional analysis, it turns out that $c_L$ is a length and $e_c$ is dimensionless because initially $\lambda$ is dimensionless, $c_B$ is a length, and $Q_A$, $Q_B$ have dimension of area. The range of variation of some of these parameters is restricted, as is discussed below. Writing the solution in a more explicit form

$$R(\varphi) = 2c/\{1 - e_c \cos\varphi + [(1 - e_c \cos\varphi)^2 + g_e]^{1/2}\} \qquad (7)$$

where $g_e \equiv 2ac = -[e_c - (c_L/c_B)][e_c - (c_B/c_L)]$. This solution has axial symmetry ($r_z$ axis), but the imposition of the boundary condition with the $R_{in}(\theta, \varphi)$ surface would in general break the axial symmetry of the final solution.

In Eq. (7) the + root was chosen to allow the important particular case $g_e = 0$ to be well defined and continuous with $g_e \neq 0$. When $g_e = 0$ the solution (7) is an ellipsoid with eccentricity $e_c$, and the main axis on the $r_z$ axis. This would require that $0 \leq e_c < 1$, and $c \geq 0$ (or $c_L/c_B \geq e_c$). The case $g_e = 0$ corresponds to two networks with the same volume characteristic, although not identical networks.

When $g_e \neq 0$, the solution is a modified ellipsoid similar to an egg form (Fig. 2) named here the g-ellipsoid (no previous identification of this surface was found). The g-ellipsoid is made by rotating the g-ellipse around the $r_z$ axis. The parameter $g_e = 2ac \geq -(1 - e_c)^2$ makes the square root of the denominator of Eq. (7) always well-defined.

The original meaning of $e_c$ as the eccentricity is formally lost when $g_e \neq 0$. In fact for $e_c > 1$ there is still a g-ellipse ($g_e > 0$) that becomes closer to a circle as $e_c$ becomes larger, as is shown in the most exterior curve of Fig 2. However, many g-ellipses are close to a single ellipse, especially when $g_e$ is small (Fig. 2). A more general approximation of the g-ellipse is obtained with two different ellipses along the $r_z$ axis. In Fig. 2. the most interior ellipse of the illustration follows a g-ellipse ($g_e = -0.109$) very closely for $r_z < 1$, similarly another half ellipse of larger eccentricity closely follows the other side of this g-ellipse ($r_z > 1.2$). This double ellipse approximation simplifies a great deal the applications of the g-ellipse discussed below. In this approximation the g-ellipse becomes a "natural interpolation" of two half ellipses of similar minor diameters with different major diameters. The reader interested in possible applications may skip to Sec. 5. The next two sections contain generalizations and additional theory.

## 3. A GENERAL ELLIPSOID AND THE ISOPERIMETRIC RELATION

So far, Eq. (6) determines the $R$ surfaces where the network volume (Eq. (3)) reach its extrema values for a given total organ volume $V_o$. It is possible to generalize Eq. (6) for any circulation distance function $X(r, \theta, \varphi)$. To write the more general equation as explicitly as possible, the required change of spherical coordinates for the network B (Fig.1) originated at $(0, 0, c_B)$, is: $r_B(r, \varphi) = (r^2 + c_B^2 - 2rc_B cos\varphi)^{1/2}$, $cos[\varphi_B(r, \varphi)] = (rcos\varphi - c_B)/r_B(r, \varphi)$, $\theta_B = \theta$. Using the general form of $X(r, \theta, \varphi)$ in Eq.(3) and Eq.(5), the extrema is now given by the generalized Eq. (6) of the $R$ surface,

$$Q_A X_A(R,\theta,\varphi) + Q_B X_B(r_B(R,\varphi),\theta,\varphi_B(R,\varphi)) = \lambda V_o. \qquad (8)$$

Which of course can not be solved in general for R, however it is clear that it has the form of a generalized ellipsoid, defined as the points where the sum of the generalized distances ($Q_A X_A$ and $Q_B X_B$) to two points ($(0,0,0)$ and $(0,0, c_B)$) is a constant ($\lambda V_o$).
The analogy to the isoperimetric theorems is established as follows. We can show also that the same Eq. (8) gives the extrema for an organ volume when a network volume value is given. Now we regard the organ volume (Eq. (4)) as a functional of $R(\theta, \varphi)$, $V = V(R)$ and assuming the only dependence of $V$ in Eq. (3) is the one explicitly written, then for a given network volume $V_{ABo} = V_{netAB}$, from Eq. (3) we have, $V(R) = I(R)/V_{ABo}$. This last equation is combined with Eq. (4) with the help of a new Lagrange multiplier $\lambda'$ in order to establish a new auxiliary functional, $V'_{aux}(R) = V(R) + \lambda'[(I(R)/V_{ABo}) - V(R)]$. This last auxiliary function can be converted to the previously defined $V_{aux}$ in Eq.(5), with the parameter identification, $V_o \lambda = V_{ABo}(\lambda' - 1)/\lambda'$, and therefore we obtain the same Eq. (8) for the surface $R(\theta, \varphi)$. The hypothesis that the extrema of the network volume is a local minima and the extrema of the organ volume is a local maxima is a more difficult problem, however this can be verified to be the case in many particular examples, including the model shown in the next section.

## 4. A FRACTAL NETWORK MODEL

Although a general derivation of Eq. (1) is available[2] it is also instructive to see how this equation approximates the volume of a more specific fractal distribution network. This Sec. presents a simple model of a class of distribution networks with fractal features of interest (Fig. 5), where the approximation given by Eq. (1) is applicable and the determination of the optimal organ form can be exactly verified.
The fractal model of this section is a simple example of a distribution network chosen primarily because the analytical methods developed so far can be computed exactly and therefore carefully compared. This fractal model example does not pretend to make a realistic representation of an organ.

There is previous work on fractal modeling of the organs[7] (see also the references in 7) exploring various optimality rules in relation to the circulatory properties of the network[8,9] a topic that this article is not concern with. In this author's opinion, the inclusion of side branching[7] provides the first realistic option in the theoretical development of the subject of fractal trees applied to the organs. However, the analytical evaluation of the global circulatory properties of a fractal network that is centrally pressurized and with side branching still remains a question open to investigation. The more realistic modeling of organ networks at the moment is being done through computer simulations, see Section 5 for further discussion on this aspect.

Next, I briefly review some of the characteristics of the distribution networks of the organs that will be desirable to include in greater detail in future developments of the subject.

It is a well-established fact that some of the main organs have distribution networks with fractal characteristics, such in the case of the lung.[7] The case of the kidney's networks is less clear in the sense of fractal design. In general terms, the networks found in organs

(*i*) have substantial side branching[7],

(*ii*) have at least three different morphological stages. The first stage is composed of the main branches of the network. Here the pulsatile flow regime is dominant and the fractal pattern is only a gross approximation. In the second stage a more typical fractal pattern is observed, and here the Poiseuille flow regime develops. In the third stage the fractal hierarchical branching of the network ends, and the basic working sites[2,3] of the organ are found, such as the nephrons and alveoli.

(*iii*) are not deterministic fractals, but are more like random hierarchical fractals that do not completely fill a given volume, and where self similar features are statistically determined[7].

(*iv*) have the ability to distribute efficiently in organ forms which are smoothly curved, that is, without the large jagged angular exterior forms found in many fractal models.

The best available mathematical treatment of a fractal network filling an arbitrary volume is found in good numerical computer simulations[4]. So far, no direct analytical procedures can be inferred from these simulations. One of the proposals of this article is that Eq. (1) can be used to find approximately the volume of fractal networks with all the characteristics listed above (*i* to *iv*) and in the likely case that some organs such as the kidney may have only minimal fractal characteristics. At this early developmental stage towards a more suitable theory of the fractal networks of organs, we still rely on deterministic and simplified fractals to test many of the ideas about these networks, which is the agenda of this Sec. in what follows.

The model of this Sec. is based on a distribution network that is a well-known deterministic fractal with side branching[7], which can only fill a simple collection of rectangles. Each rectangle is subdivided in four equal rectangles, and the branches spread diagonally from the center to the center of the smaller rectangles. This fractal branching pattern is shown in Fig. 3, where it is applied to the 10 rectangles near the top and side exterior border. The overall volume to be serviced by the network is a rectangular volume with sides $a$ and $b$, with a very small and constant thickness $h$. To resemble a kidney form, the ($a$)($b$) rectangle has a ($a/3$)($b/3$) rectangular indentation at the entering region of the main branches of the network. The region where this simulated blood supply network develops fractal branching, here represented by the layer of 10 rectangles each with area

= $(b/3)(a/6)$ in Fig. 3, is chosen to simulate the nephron region that is near the exterior surface. The $A_k$ are the cross sections of the branches and the single main branch with $A_0 = 2A_1$ is partially shown in Fig. 3. This fractal model is two-dimensional and although it could be made three-dimensional[7], this more complicated option would not yield significant additional insights at this time.

This model has the features described in (*i*) and (*ii*), and some aspects of (*iii*). There are many choices in the selection of the location of the first and second order branches of the network. The one shown in Fig. 3 was selected because of the good correlation with Eq. (1) as is shown next. The distribution network volume in Fig. 3 can be evaluated exactly and directly as

$$V_{nete} = [4A_1 + 9A_2 + n_f(3A_3/4 + 11A_4/8 + 43A_5/16 + \ldots + c_s A_s/2^{s-1})][(a/6)^2 + (b/3)^2]^{1/2}$$
$$= F(A_k)[(a/2)^2 + b^2]^{1/2} \qquad (9)$$

where $n_f = 10$ and the numerical coefficients $c_k$ follow the limit ratio $c_{k+1}/c_k \longrightarrow 4$. The given number $s$ labels the last branching stage that services the sites. A site volume = $h(a/6)(b/3)/4^{s-2}$ = nephron volume of this rectangular kidney. As it is evident in Eq. (9) the network volume separates into two factors, one containing the organ form with parameters $a$ and $b$, and the other factor $F(A_k)$ containing the cross section areas $A_k$. There is an arbitrary constant numerical factor in the choice of $F(A_k)$. Notice that other combinations of $(a/6)(b/3)$ rectangles can be set to have fractal branching and be accounted for in Eq. (9) by simply changing $n_f$. We could selectively have a different number of initial branching generations without self-similar branching. The value of $n_f$ could be modified randomly and/or due to boundary conditions requiring the exclusion of some rectangles, in order to accommodate other organs. These are all the possible networks of variable fractal characteristics that are accounted for in this model. All the networks under consideration have a $F(A_k)$ independent of the $a$, $b$ parameters, or with a dependence that has the same effect on the minimization results, such as the likely possibility that $F(A_k)$ is a function of $abh$. Therefore the $A_k$ need not be determined in this analysis and can be kept as a separate problem from the optimal organ form. The determination of the minimum of $V_{nete}$ in relation to the $a, b$ parameters, while keeping $V$ constant, is formulated with a Lagrange multiplier $L_g$ and the auxiliary function

$$V_{aux1} = F(A_k)[(a/2)^2 + b^2]^{1/2} + L_g(V - (8/9)abh), \qquad (10)$$

and by setting the equations: $\partial V_{aux1}/\partial a = 0$; $\partial V_{aux1}/\partial b = 0$. The solution of these equations is simple and it is $a/b = 2$. This solution determines the optimal dimensions of the rectangular organ form.

Now let us examine the general approximation proposed by Eq. (1) to the above network model shown in Fig. 3,

$$V_{net} = PI_X/V = PV^{-1}hf_1 \left( \int_{-a/2}^{a/2} \int_0^b - \int_{-a/6}^{a/6} \int_0^{b/3} \right)(x^2 + y^2)^{1/2} dx dy \qquad (11)$$

with $V = 8abh/9$. In Eq. (11) the first proposed dominant term of the circulation distance function $X(\mathbf{r}) = f_1 r = f_1 (x^2 + y^2)^{1/2}$ is already introduced where $f_1$ a constant in relation to the coordinates. Notice that Eq. (1) contains a general separate factor for the organ form parameters, which is $V^{-1}I_X = V^{-1}\int_{V'} X(\mathbf{r})dV$. The other factor $P$ contains the cross section variables ($A_k$). This $P$ factor can be equated with the factor $F(A_k)$ of Eq. (9), for a given selection of the $A_k$. The integrals of Eq. (11) can be evaluated exactly in terms of the following function $G_0(u)$,

$$3G_0(u) \equiv 6\int_0^1\int_0^1 (u^2x^2 + y^2)^{1/2} dxdy = 2(1+u^2)^{1/2} + u^{-1}\text{arcsinh}(u) + u^2\text{arcsinh}(u^{-1}). \quad (12)$$

The inverse hyperbolic function is defined as $\text{arcsinh}(u) = \ln(u + (u^2+1)^{1/2})$
$= \int_0^u (1+z^2)^{-1/2} dz$. Since $G_0(u=0) = 1$, there is not a true singularity at $u = 0$ in Eq. (12). The $G_0(u)$ function has the useful symmetry $G_0(u) = uG_0(1/u)$. By using Eq. (12) the integrals of Eq. (11) are evaluated as follows,

$$\frac{V_{net}}{P} = \frac{hf_1}{V}(1-3^{-3})\frac{ab^2}{2}G_0(\frac{a}{2b}) = \frac{13}{24}f_1 bG_0(\frac{a}{2b}). \quad (13)$$

The numerical factor $f_1$ will be determined later, however it has no effect on the optimal values of the $a$, $b$ parameters. To determine these optimal values as before, an auxiliary function $V_{aux2}$ with a Lagrange $L_g$ multiplier is defined as $V_{aux2} = F_2 bG_0(a/2b) + L_g(V - (8/9)abh)$, where $F_2 = Pf_1(13/24)$ are all the other factors of the network volume that do not depend on $a$ or $b$ or depend only on $ab$. Then the equations are
$\partial V_{aux2}/\partial a = 0; \partial V_{aux2}/\partial b = 0$. After simplifying these equations the resulting equation is $G_0(v) = 2vdG_0(v)/dv$, with $v = a/2b$. The solution of this last equation is $v = a/2b = 1$, which remarkably, is the same solution previously found for $V_{aux1}$ from Eq. (10).
The last step is to compare the exact network volume given in Eq. (9) and the result given by Eq. (13). The constant $f_1$ is set by requiring that both formulas for the network volume give the same result for $v = a/2b = 1$, which makes $f_1 = 2^{1/2}24/13G_0(1) = 1.706$. A good way of comparing the network volumes given by Eq. (9) and Eq. (13) is to look at the ratio $V_{nete}/V_{net} = [(a/2)^2 + b^2]^{1/2}G_0(1)/2^{1/2}bG_0(a/2b) = [v^2 + 1]^{1/2} / 0.9421G_0(v)$, which depends only on the ratio $v = a/2b$. This is shown in Fig. 4 where we see that the approximation given by Eq. (1) to the network volume of the model shown is good, even for other than the optimal values ($a = 2b$) of the organ volume parameters.

## 5. THE ORGAN FORM MODELING ISSUES

With the help of computer technology, it is currently possible to build almost realistic distribution networks with fractal characteristics filling 3-dimensional organ-like volumes. For a review of the CCO method, one of the computer techniques in use, and its results, the interested reader is referred to Schreiner[4,10] *et al.* and for more computer assisted modeling, to Schreiner[11] *et al.* and Schmidt[12] *et al.*
Several of the results described by Schreiner[4,10] *et al.* lend support to the model proposed in this article. For example, quoting a paragraph from Schreiner[10], from the section

"Shape of perfusion domain" (note that "perfusion domain" correspond here to organ form and "intravascular volume" is network volume):

"Variations in the shape of perfusion domain as well as variations in the site of inlet lead to different branching patterns, see figure 11. When optimizing according to minimum intravascular volume, different shapes and/or sites of inlet are seen to require amounts of blood differing by a factor of two (i.e. costly 'dead volume') in order to supply all sites of 2-dimensional perfusion domains (35). In general, in terms of the value of the target function (total intravascular volume), lengthy shaped organs are perfused 'less efficiently' than compact ones, and placing the inlet at the 'long side' of an organ is more efficient than at the 'short side'."

These conclusions are consistent with the results of the model presented in this article. A general finding here is that the minimization of network volume in relation to organ form is, to a first approximation, independent of the requirements of efficient circulation, which is also confirmed by Schreiner[4,10] *et al*. There are several areas of convergence between those computer simulations[4,10,11,12] and this article that would be worth further investigation.

To have a model for $R(\theta, \varphi)$ even closer to real organ forms some other elements from condition (*III*) must be included. In the absence of detailed information on the organ networks characteristics there are the following options to investigate. One route for a more general modeling method of the organ form may be found by introducing more general forms of the circulation distance, $X(r, \theta, \varphi) = f_1 r + f_2 r^v \chi(\theta, \varphi)$ as test functions and proceed with Eq. (8) to determine the $R(\theta, \varphi)$ surfaces. Another route is to consider the asymmetric restrictions on the available space to the organ along the *x, y,* and *z* directions. We can incorporate this in the arguments leading to Eq. (7) by considering only an angular slice, between $\theta$ and $\theta + \Delta\theta$, of the total organ volume as the region of integration. Then the parameters $\lambda$ and $V_o$ are replaced by functions of $\theta$. The new solution for $R$ is still given by Eq. (7), with new parameters $c$, $e_c$ and $g_e$ which are functions of $\theta$, which in turn depend on the way the additional boundary conditions are imposed.

These arguments show that a more complete solution for $R$ can be found within a larger family of surfaces generated by g-ellipses. Such solution is constructed with a more general g-ellipsoid that has, in the double ellipsoid approximation, different eccentricities along the *x, y, z* directions. Such surface is here called the g3-ellipsoid.

The use of the g3-ellipsoid and ellipsoid in modeling the kidney and the overall exterior form of both lungs is shown in Fig. 5 and 6. The fitness of the g-ellipse is observable by looking at the cross sections and at the two-dimensional projections of the organ forms. Other exterior views of the kidney (not shown) display a similar fitness to the g-ellipse, as seen in Fig. 5. These illustrations are not intended as rigorous evaluations, but rather as examples of possible applications of the model developed so far. No statistical analysis of these results has been attempted, an issue that by itself would require further theoretical development.

There is plenty of medical imaging of the lung of normal human subjects, where the type of curve fitting shown in Fig. 6 is a very common result, but the traditional anatomical drawing of Fig. 6 is a much clearer illustration. Fig. 6 suggests that the exterior form of both lungs follow a single optimal organ design given by a section of an ellipsoid or g3-ellipsoid. It would be very interesting to explore further this description of the lungs at the various stages of the breathing cycle.

The surfaces generated by the ellipse and the g-ellipse capture only one side of the organ form. On its other side, there is a more complicated surface that is designated by $R_{in}(\theta, \varphi)$ (Eq. (4)). In the kidney this surface is in the region where the main blood vessels and ureter are found, and in the lung, we find the diaphragm and other internal organs. Remarkably, a very general surface $R_{in}(\theta, \varphi)$ can coexist with the g-ellipsoid and be in full compliance of the stated isoperimetric optimality principle. The modeling of the surface $R_{in}(\theta, \varphi)$ which in this theory (condition (*III*)) is mainly determined by the spatial coexistence of the surrounding internal organs remains an important and arduous task. These, and other modeling issues about the organs are now available to quantitative analysis through the possible applications and extensions of this model.

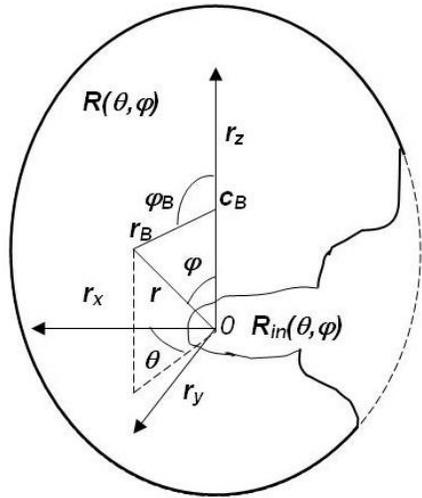

FIG. 1. Schematics of coordinates and geometric variables of model. (Sec. 2)

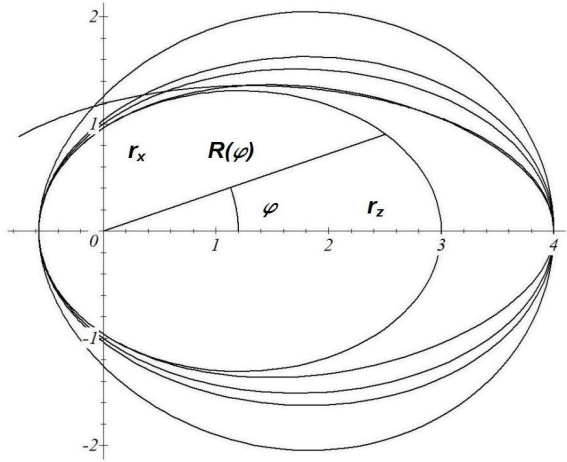

FIG. 2. Ellipses ($g_e = 0$), and g-ellipses compared. From Eq. (7), the parameters of the five closed curves from inside to outside are:
$c$ = 0.96 ; 0.933; 1.0 ; 1.07 ; 1.833
$e_c$ = 0.68 ; 0.65 ; 0.75 ; 0.85 ; 2.0
$g_e$ = 0.0 ; –0.109; 0.0 ; 0.124; 2.674
The cut-off curve ($r_x > 0$) is the ellipse: $(r_x/1.355)^2 + ((r_z-1.3)/2.7)^2 = 1$; (ecc. = 0.865).
This cut-off ellipse together with the most interior ellipse ($g_e = 0.0$) shows the double-ellipse approximation to the g-ellipse with $g_e = -0.109$

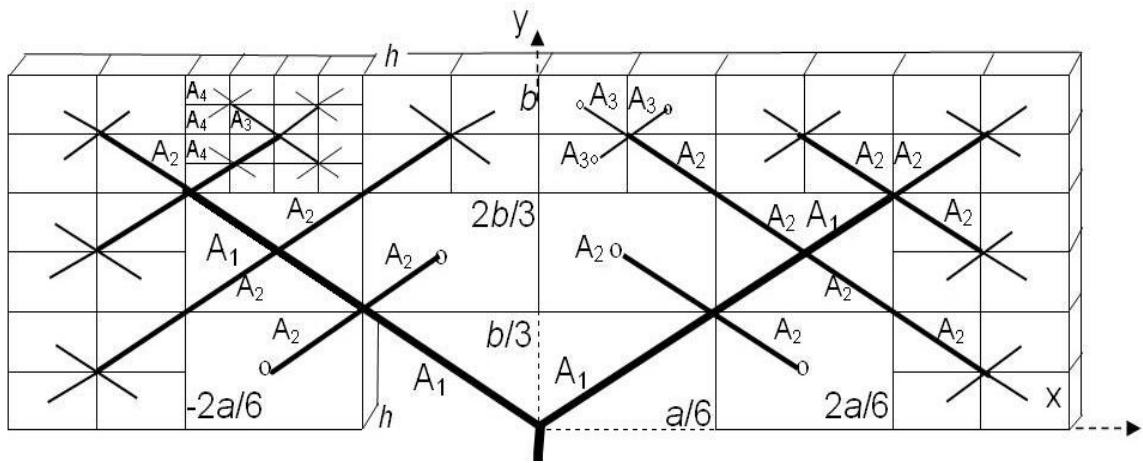

FIG. 3. Model discussed in Sec. 4 of a fractal network servicing a "rectangular organ" form. Fractal branching in the 10 exterior rectangles each having area ($a/6$)($b/3$). For the sake of clarity only one of these rectangles has been illustrated with a fourth order branching.

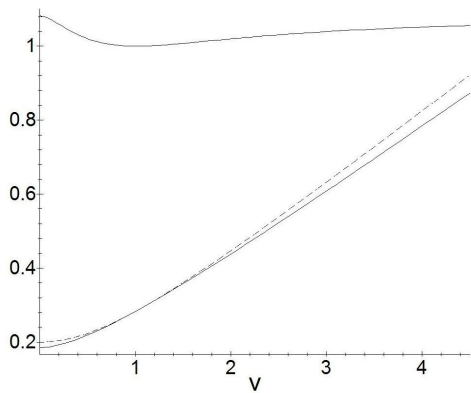

FIG. 4. From the top, the graphs of:
$(1 + v^2)^{1/2}/0.9421 G_0(v)$, $(1 + v^2)^{1/2}/5$,
$0.9421 G_0(v)/5$, from Sec. 4. The
function $G_0(v)$ is given by Eq. (12).
The (1/5) factor is included in order to
show all graphs in the same axis scales.

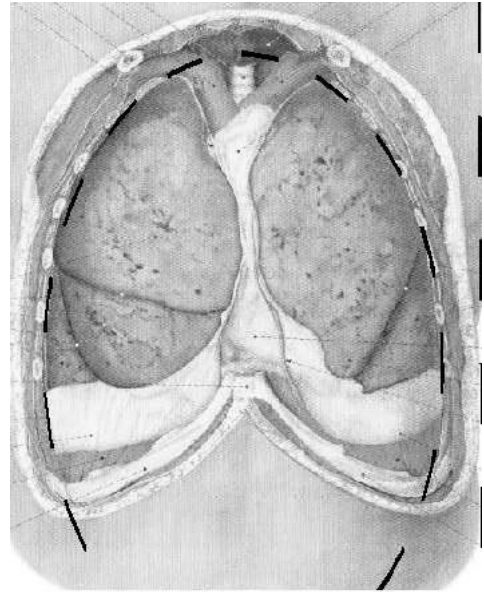

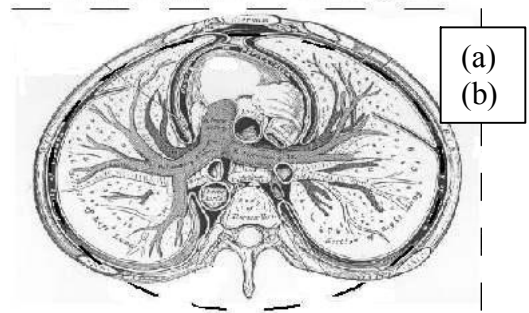

(a)
(b)

FIG. 6. Human lung forms approximated
by an ellipsoid (dashed curves). Part (a),
in situ illustration[5] of thorax. Part (b) is a
transverse section[6] of the thorax shown
in (a) at the level of the heart. See Sec. 5.

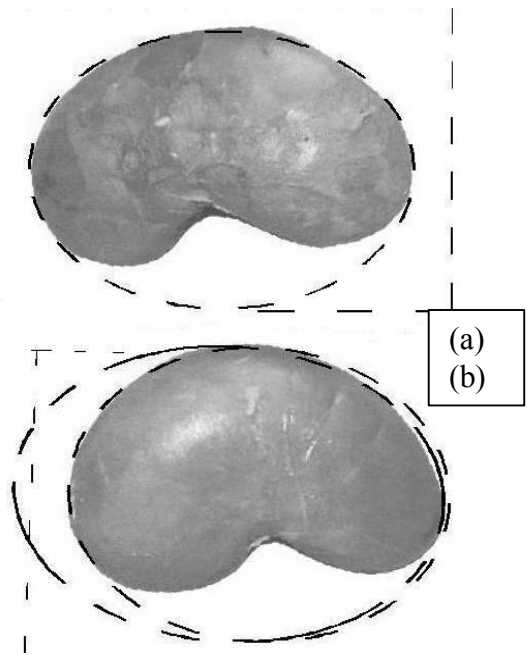

(a)
(b)

FIG. 5. Photographs of sheep's kidneys,
approximated by an ellipse (a), and by
a g-ellipse (b) using the method of the
double ellipse. The interior common area
of both ellipses in (b) is the area defined
by the g-ellipse. The dashed straight
lines show the axis directions of the
ellipses. See Sec. 5.

...\MIOF\\M3G from M3F

12